\documentclass[12pt,preprint]{aastex}

\slugcomment{ver. 2009.07.29}

\shorttitle{Polarized flares from Sgr A*}
\shortauthors{Nishiyama et al.}

\begin{document}

\title{Near-infrared Polarimetry of flares from Sgr A* with Subaru/CIAO
}

\author{Shogo Nishiyama\altaffilmark{1,2}, 
Motohide Tamura\altaffilmark{3},
Hirofumi Hatano\altaffilmark{4}, 
Tetsuya Nagata\altaffilmark{1},
Tomoyuki Kudo\altaffilmark{3},
Miki Ishii\altaffilmark{5},
Rainer Sch\"{o}del\altaffilmark{6},
and Andreas Eckart\altaffilmark{7,8}
}

\altaffiltext{1}{Department of Astronomy, Kyoto University, 
Kyoto, 606-8502 Japan; shogo@kusastro.kyoto-u.ac.jp}

\altaffiltext{2}{Research Fellow of the Japan Society for the Promotion of Science (JSPS)}

\altaffiltext{3}{National Astronomical Observatory of Japan, 
Mitaka, Tokyo, 181-8588 Japan}

\altaffiltext{4}{Department of Astrophysics, Nagoya University, 
Nagoya, 464-8602 Japan}

\altaffiltext{5}{Subaru Telescope, National Astronomical Observatory of Japan, 650 
North A`ohoku Place, Hilo, HI 96720, USA}

\altaffiltext{6}{Instituto de Astrof\'{i}sica de Andaluc\'{i}a (IAA)-CSIC, Camino Bajo de Hu\'{e}tor 50, E-18008 Granada, Spain}

\altaffiltext{7}{I.Physikalisches Institut, Universit\"{a}t zu K\"{o}ln, Z\"{u}lpicher Str. 77, 50937 K\"{o}ln, Germany}

\altaffiltext{8}{Max Planck Institute f\"ur Radioastronomie, Auf dem H\"ugel 69, 53121 Bonn, Germany}

\begin{abstract}

We have performed near-infrared monitoring observations of Sgr A*, 
the Galactic center radio source associated with a supermassive black hole,
with the near-infrared camera CIAO and the 36-element adaptive optics system on the Subaru telescope.
We observed three flares in the $K_S$ band (2.15$\mu$m) during 220 minutes monitoring on 2008 May 28,
and confirmed the flare emission is highly polarized, supporting
the synchrotron radiation nature of the near-infrared emission.
Clear variations in the degree and position angle of polarization were also
detected: an increase of the degree of polarization of about 20 \%, 
and a swing of the position angle of $\sim 60\degr - 70\degr$ in the declining phase of the flares. 
The correlation between the flux and the degree of polarization 
can be well explained by the flare emission coming from hotspot(s) orbiting Sgr A*.
Comparison with calculations in the literature gives 
a constraint to the inclination angle $i$ of 
the orbit of the hotspot around Sgr A*, as $45\degr \leq i<90\degr$ (close to edge-on).

\end{abstract}


\section{Introduction}
\label{sec:intro}

Sagittarius A* (Sgr A*), the compact source of radio, infrared, and X-ray emission
at the center of our Galaxy, is associated with a supermassive black hole \citep[SMBH;][]{Eckart96,Ghez00,Schodel02}
weighing in at $\approx 4 \times 10^6 M_{\sun}$ \citep{Ghez08,Gillessen09}.
Since the first identification of Sgr A* in the near-infrared (NIR) wavelengths \citep{Genzel03},
NIR flares have been studied extensively.
Such observations have found that Sgr A* is highly variable,
and periodicity of $\sim 20$ minutes claimed in monitoring observations \citep[e.g.,][]{Genzel03,Eckart06}
would place the corresponding emission region at roughly three Schwarzschild radii
for the black hole mass.
If this picture is correct, 
the NIR flares from Sgr A* are powerful tools that allow us 
to investigate an accretion flow and the behavior of matter and radiation 
in the strong gravity regime very close to the SMBH.

Polarization in the NIR bands stands as one of the few observables of 
Sgr A* not extensively investigated due to its difficulties \citep{Meyer06b}.  
Successful NIR polarimetry of Sgr A* needs 
(i) the occurrence of a fairly bright flare, 
(ii) quick measurements to trace the intrinsic rapid variation of polarization,
and (iii) excellent and stable atmospheric conditions during the entire observations.
Up to now, more than a dozen NIR flares have been observed photometrically, 
but only four have been observed in linear polarization 
\citep{Eckart06,Meyer06a,Meyer06b,Eckart08}, 
with all the observations carried out 
with the NAOS/CONICA at the Very Large Telescope (VLT).  
Also, the duration of the observations was generally not very long (about 100 minutes).  
Time-resolved polarized light curves carry
specific information about the interplay between the gravitational and magnetic fields
near the Sgr A* event horizon, because the propagation of the polarization vector
is sensitive to the presence and properties of these fields along the light trajectories \citep{Genzel07}.

In this letter, we present the first independent polarimetric measurements
using the NIR camera CIAO and the adaptive optics (AO) system AO36
on the 8.2 m telescope Subaru\footnote{
Based on data collected at Subaru Telescope, 
which is operated by the National Astronomical Observatory of Japan}.
This is an almost non-stop polarimetric observation of Sgr A* over 220 minutes, 
and we have detected three main flares and shorter sub-flares
superimposed on the underlying main flares.
We will show that the profile is fairly well explained 
in the context of a hotspot orbiting the SMBH.

\section{Observations and Data Analysis}
\label{sec:obs}

\begin{figure}[h]
 \begin{center}
  \epsscale{.75}
    \plotone{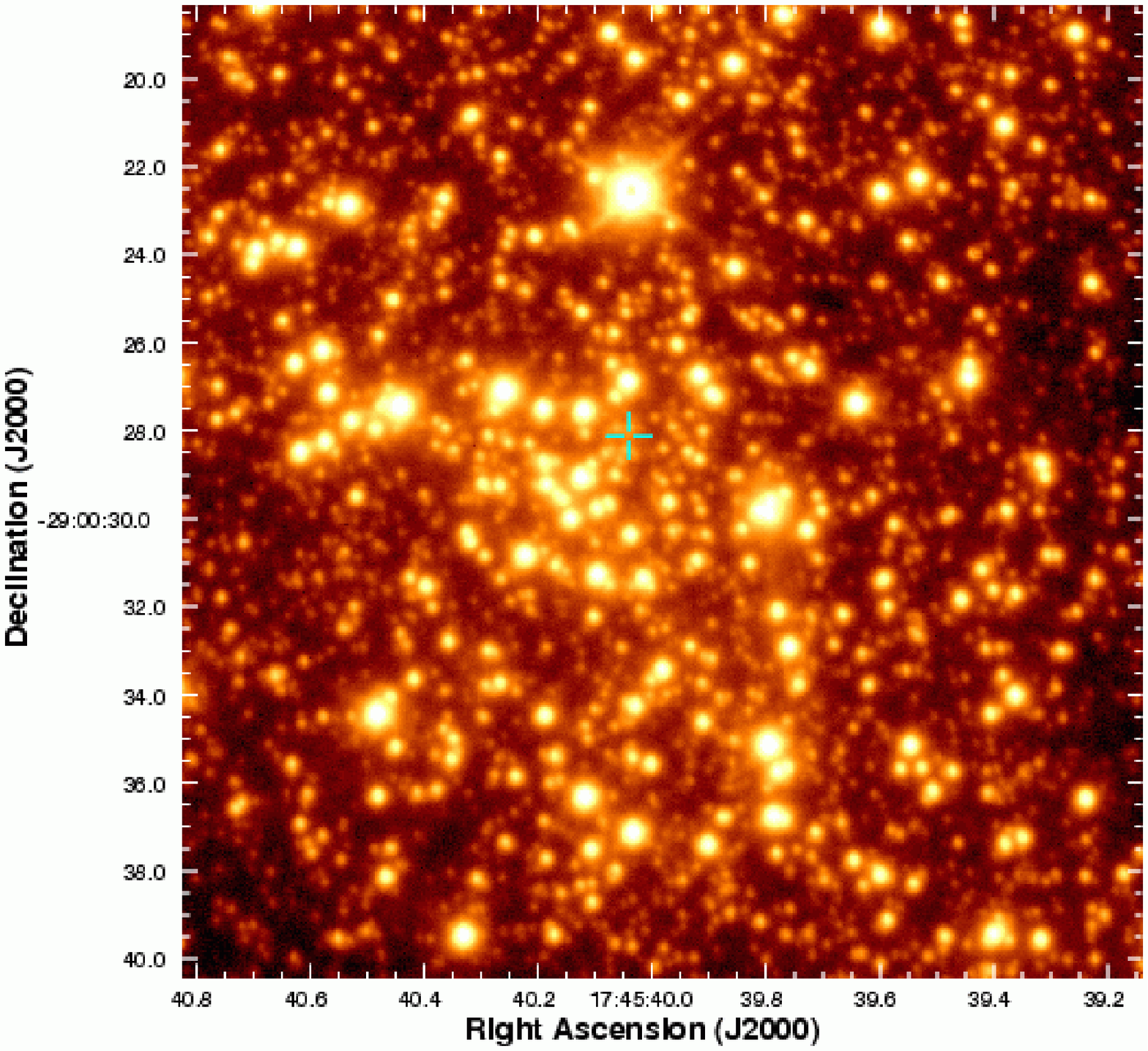}
    \caption{
      $K_S$-band AO image of the central $22\arcsec$ of our Galaxy,
      obtained with the NIR camera CIAO on the Subaru telescope.
      The position of Sgr A* is indicated by the cyan lines.
      The intensity scale is logarithmic, and the integration time is 20 sec.
    }
  \label{fig:GCImage}
 \end{center}
\end{figure}

On 2008 May 28, we conducted $K_S$ band ($2.15 \mu$m) polarimetric observations of Sgr A*
using CIAO and its polarimeter \citep{Tamura03} and AO36 \citep{Takami04}
on the Subaru telescope \citep{Iye04}.
CIAO provides an image of a $22\farcs2 \times 22\farcs2$ area of sky
with a scale of 21.7 mas pixel$^{-1}$ (Fig. \ref{fig:GCImage}).
With the $R = 13.2$ mag natural guide star USNO 0600-28577051
located about $30\arcsec$ from Sgr A*,
and stable atmospheric condition during the observations,
AO36 provided a stable correction with seeing values 
between 0\farcs17 and 0\farcs21 in the $K_S$ band.
CIAO employs a rotatable half wave plate installed in front of the AO
and a fixed wire grid polarizer in the cryostat to measure linear polarization.
We made 20 sec exposures at four wave plate angles
in the sequence of $0\fdg0$, $45\fdg0$, $22\fdg5$, and $67\fdg5$.\footnote{
In this sequence of observation, rapid variation of Sgr A* flux 
can bias polarization measurements, 
but we confirmed that such ``instrumental polarization'' is 
only a few \% except when the flux changes rapidly at 150 minutes 
in the current observation (Fig. \ref{fig:LCSgrASStars}).  
}
Including time for readout and rotation of the wave plate,
about 200 sec is required for one sequence of observation.
We observed a dark cloud located at a few arcmin northwest from Sgr A*
before and after the observations of Sgr A* to obtain sky measurements.

\begin{figure}[h]
  \begin{minipage}{0.45\linewidth}
    \begin{center}
      \epsscale{1.0}
      \plotone{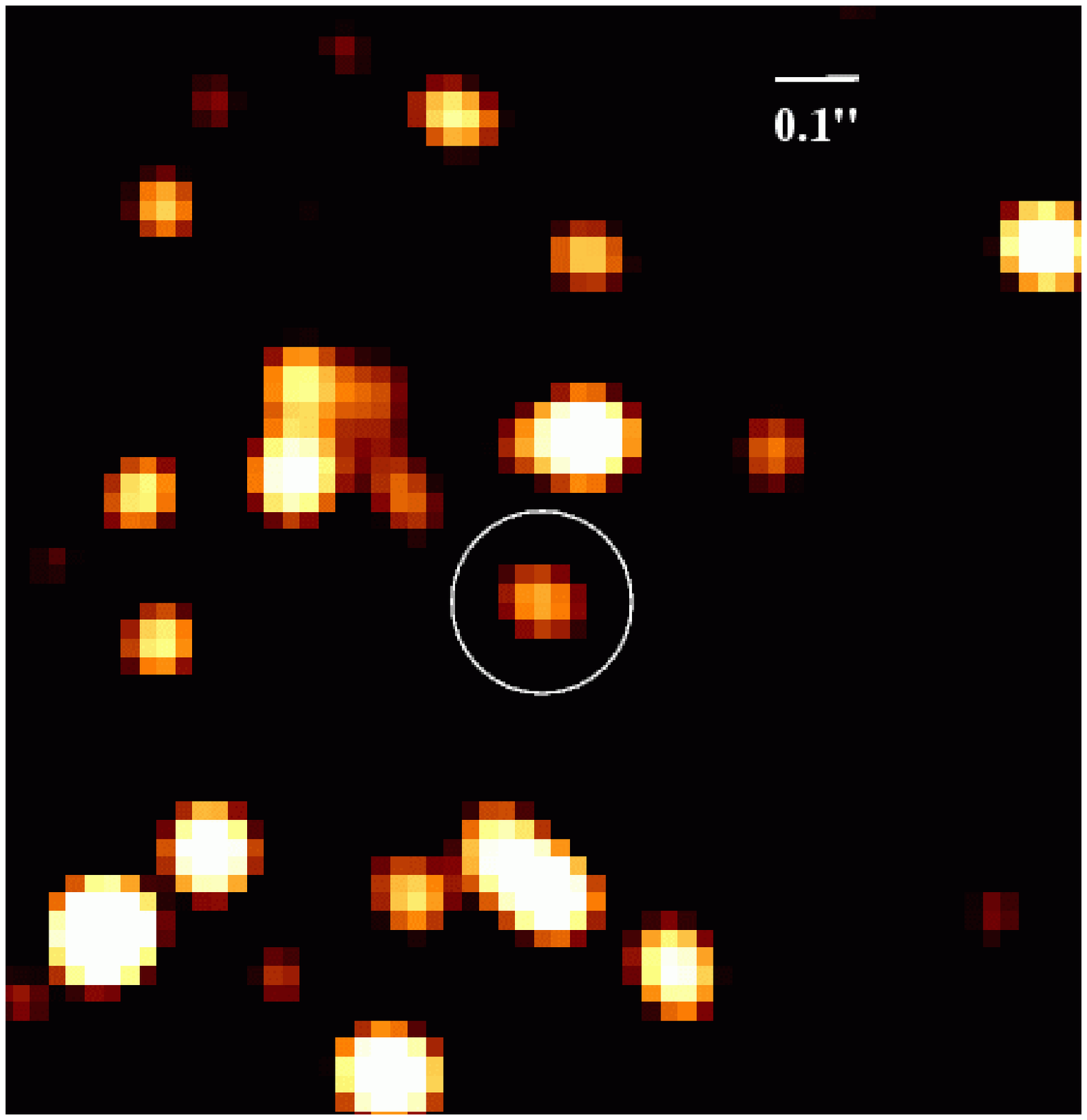}
    \end{center}
  \end{minipage}
  \hspace{0.05\linewidth}
  \begin{minipage}{0.45\linewidth}
    \begin{center}
      \epsscale{1.0}
      \plotone{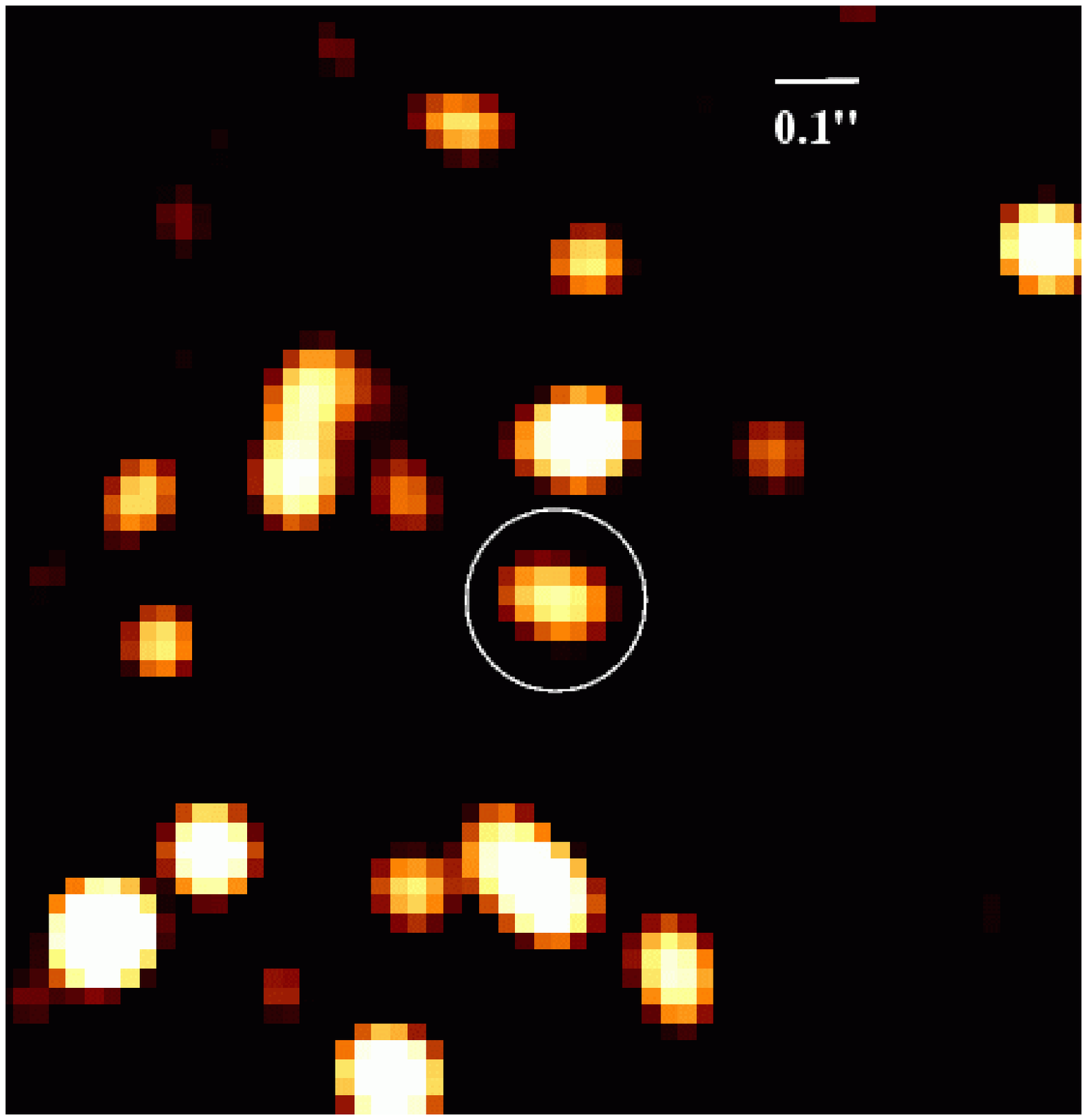}
    \end{center}
  \end{minipage}
  \caption{
    Central $1\farcs3 \times 1\farcs3$ region of the GC images in the $K_S$ band.
    The time from the beginning of the observation is 145.1 (left) and 158.4 min (right).
    In each image, the location of Sgr A* is marked by a circle,
    and a faint star (S17) is also located in the circle.
    The image scale is logarithmic, and the integration time is $20 \times 4$ sec.
    The images were sky-subtracted, flat fielded, corrected for bad pixels and cosmic rays,
    and deconvolved with the Lucy-Richardson algorithm.
  }
  \label{fig:ZoomImages}
\end{figure}

IRAF (Image Reduction and Analysis Facility)\footnote{
IRAF is distributed by the National Optical Astronomy
Observatory, which is operated by the Association of Universities for
Research in Astronomy, Inc., under cooperative agreement with
the National Science Foundation.}
software package was used
to perform dark- and flat-field corrections, sky background estimation and subtraction
followed by bad pixel corrections.
A point spread function was extracted via point-source fitting 
with the program {\it StarFinder} \citep{Diolaiti00} from each image.
Each image was then deconvolved with the Lucy-Richardson algorithm \citep{Lucy74,Richardson72},
and beam restored with a Gaussian beam of FWHM corresponding to
the respective wavelengths (66 mas at 2.15 $\mu$m).
The deconvolved images show clear flares at the position of Sgr A* (Fig. \ref{fig:ZoomImages}).

Flux densities of Sgr A* and other point sources were obtained via aperture photometry
in Stokes $I$ images [$I=(I_{0\degr}+I_{22\fdg5}+I_{45\degr}+I_{67\fdg5})/2$].
We used DAOFIND and APPHOT tasks for point sources identification and photometry.
The flux density of each $I$ image was calibrated with 
10 sources in the same field.
For extinction correction, we assumed $A_{K_S} = 2.8$ mag \citep{Eisenhauer05}.
The flux was corrected for a background flux density contribution $F_{\mathrm{bg}}$,
which was determined as the mean flux measured in the same aperture size at six different positions,
which are free from contributions of individual stars.
The photometric error at each flux of Sgr A* was estimated
by fitting a power-law to the rms uncertainty 
in the flux for non-variable faint stars of similar flux densities.
We found a typical dependence of the photometric error $\sigma$ on flux $F$ as
$\sigma = 0.06F^{0.78}$ mJy.

The Stokes parameters $I, Q$, and $U$ for Sgr A* and other point sources were
determined by aperture polarimetry as follows.
DAOFIND and APPHOT tasks were used to obtain an intensity for each wave plate angle
($I_{0\degr}$, $I_{22\fdg5}$, $I_{45\degr}$, $I_{67\fdg5}$)
with an aperture radius of 5 pixel (0\farcs11, see a circle in Fig. \ref{fig:ZoomImages}).
Based on the intensities, we calculated the Stokes parameters as
$I = (I_{0\degr} + I_{22\fdg5} + I_{45\degr} + I_{67\fdg5})/2$,
$Q = I_{0\degr} - I_{45\degr}$, and $U = I_{22\fdg5} - I_{67\fdg5}$.
With these Stokes parameters, 
the degrees $P$ and position angles $\theta$ of polarization for stars in the same field were obtained.
We confirmed that $P = 10.6 \%$ and $\theta = 19.8 \degr$ for IRS 21,
and $4.8 \%$ and $24\degr$ for mean polarization of field stars;
these agree well with the results by \citet{Ott99}.
We obtained a small mean RMS of the degrees of polarization 
as $0.9 \%$ for the non-variable faint sources in the same field,
suggesting a stable atmospheric condition and a reliable photometry.

To obtain intrinsic degree and position angle of polarization for Sgr A*,
contributions of interstellar polarization and a faint nearby star S17 should be taken into account.
Here we assume that S17 is an intrinsically unpolarized source.
An observed Stokes parameter normalized by intensity, ${\left( \frac{Q}{I} \right)}_{\mathrm{obs}}$,
is given by
\begin{eqnarray}
  {\left( \frac{Q}{I} \right)}_{\mathrm{obs}} \approx {\left( \frac{Q_{\mathrm{Sgr A^*}}}{I_{\mathrm{Sgr A^*}}+I_{\mathrm{S17}}} \right)}
  + {\left( \frac{Q}{I} \right)}_{\mathrm{ISM}}, 
\label{eq:QIUIObs}
\end{eqnarray}
where $Q_{\mathrm{Sgr A^*}}$ is a Stokes $Q$ parameter for Sgr A*,
$I_{\mathrm{Sgr A^*}}$ and $I_{\mathrm{S17}}$ are Stokes $I$ parameters for Sgr A* and S17, respectively,
and ${\left( \frac{Q}{I} \right)}_{\mathrm{ISM}}$ is a normalized Stokes $Q$ parameter for the interstellar polarization.
Hence, we can obtain an intrinsic, normalized Stokes $Q$ parameter for Sgr A* as
\begin{eqnarray}
  {\left( \frac{Q_{\mathrm{Sgr A^*}}}{I_{\mathrm{Sgr A^*}}} \right)} \approx \Bigg[ {\left( \frac{Q}{I} \right)}_{\mathrm{obs}} - {\left( \frac{Q}{I} \right)}_{\mathrm{ISM}} \Bigg] 
  \times {\left( \frac{I_{\mathrm{Sgr A^*}} + I_{\mathrm{S17}}}{I_{\mathrm{Sgr A^*}}} \right)}.
\label{eq:QIUISgrASInt}
\end{eqnarray}
${\left( \frac{Q}{I} \right)}_{\mathrm{ISM}}$ (and ${\left( \frac{U}{I} \right)}_{\mathrm{ISM}}$) can be calculated accurately with bright stars in the same fields,
as ${\left( \frac{Q}{I} \right)}_{\mathrm{ISM}} = 0.032$ and 
${\left( \frac{U}{I} \right)}_{\mathrm{ISM}} = 0.022$,
in the assumption that they are located near the Galactic Center and are intrinsically unpolarized.
We derived the last term in Eq. (\ref{eq:QIUISgrASInt}) as
$(I_{\mathrm{Sgr A^*}} + I_{\mathrm{S17}})/(I_{\mathrm{Sgr A^*}}) 
= (F_{\mathrm{Sgr A* + S17}}-F_{\mathrm{bg}})/(F_{\mathrm{Sgr A* + S17}}-(F_{\mathrm{S17}}+F_{\mathrm{bg}}))$,
where $F_{\mathrm{Sgr A* + S17}}$ is the total flux density of Sgr A* and S17,
$F_{\mathrm{bg}}$ is a background flux density contribution,
and $F_{\mathrm{S17}} = 6.7$ mJy  \citep[dereddened flux; an observed magnitude is 15.3 mag, see][]{Gillessen09}.
With the parameters $(Q_{\mathrm{Sgr A^*}}/I_{\mathrm{Sgr A^*}})$ 
and $(U_{\mathrm{Sgr A^*}}/I_{\mathrm{Sgr A^*}})$
and equations 
\begin{eqnarray}
P_{\mathrm{Sgr A^*}} = \sqrt{{\left( \frac{Q_{\mathrm{Sgr A^*}}}{I_{\mathrm{Sgr A^*}}} \right)}^2 
  + {\left( \frac{U_{\mathrm{Sgr A^*}}}{I_{\mathrm{Sgr A^*}}} \right)}^2},
\theta_{\mathrm{Sgr A^*}} = \frac{1}{2} \arctan \Bigg[ {\left( \frac{U_{\mathrm{Sgr A^*}}}{I_{\mathrm{Sgr A^*}}} \right)} \Bigg/ {\left( \frac{Q_{\mathrm{Sgr A^*}}}{I_{\mathrm{Sgr A^*}}} \right)}\Bigg],
\label{eq:PThetaSgrAS}
\end{eqnarray}
we can obtain the intrinsic degree $P_{\mathrm{Sgr A^*}}$ 
and the position angle $\theta_{\mathrm{Sgr A^*}}$ of polarization for Sgr A*. 
The position angle $\theta$ is measured from the north and increasing counterclockwise.

\section{Results}
\label{sec:Results}

\begin{figure}[h]
 \begin{center}
  \epsscale{.80}
    \plotone{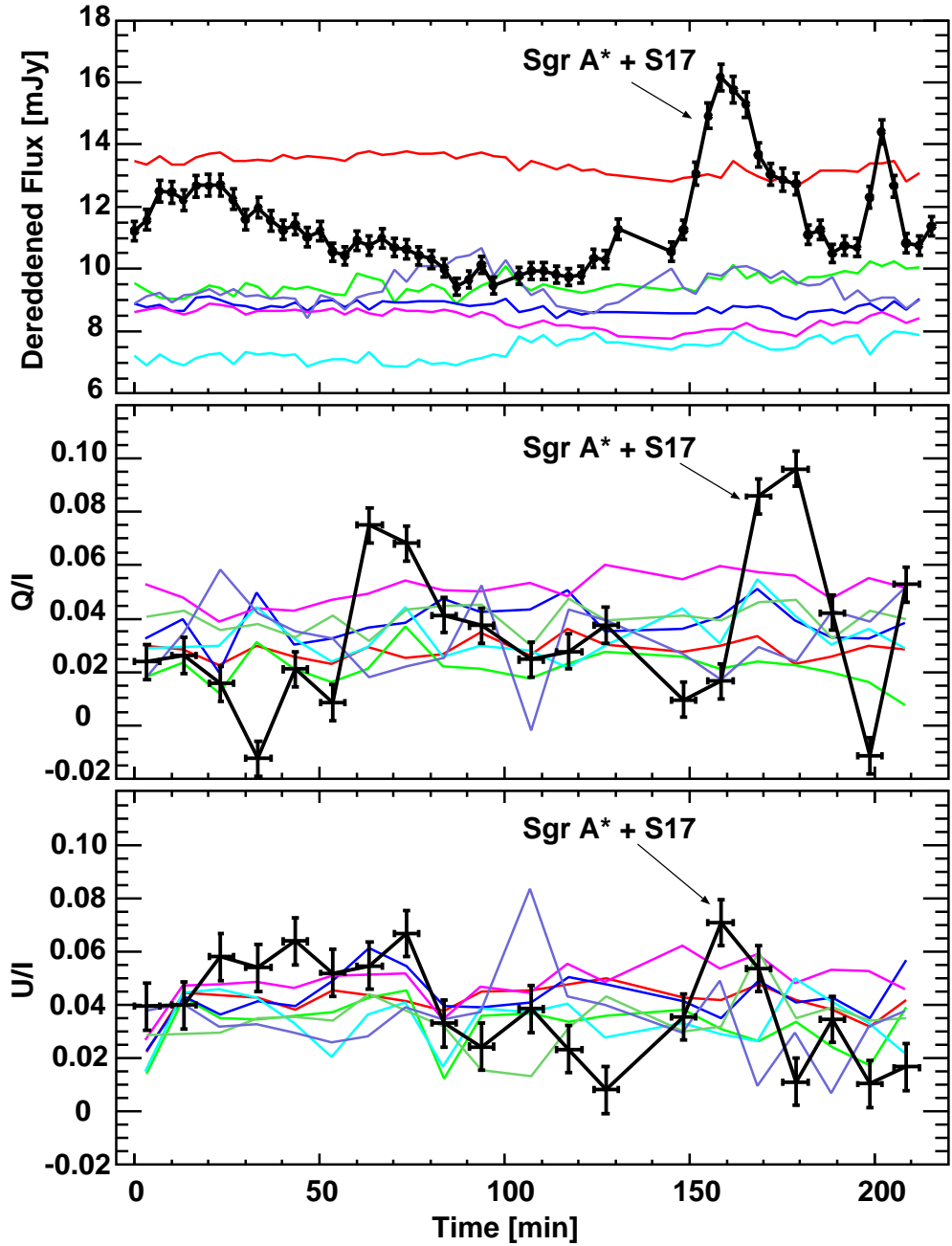}
    \caption{
      Top: light curves of Sgr A* plus S17 flux (thick black line), 
      and other faint stars in the same field (colored thin lines).
      Middle and bottom: time evolution of $Q/I$ (middle) and $U/I$ (bottom)
      for Sgr A* plus S17 flux (thick black line), 
      and other faint stars in the same field (colored thin lines).
    }
  \label{fig:LCSgrASStars}
 \end{center}
\end{figure}

The dereddened flux of Sgr A* plus S17 show a clear variation 
(Fig. \ref{fig:LCSgrASStars}, top panel, black line).
For reference, 
those for the non-variable faint sources in the same field are also shown in the same panel (colored lines).
There are three major peaks:
the broadest, and the weakest flare from $t = 0$ to 90 minutes,
the strongest flare from 145 to 190 minutes, in which a peak is followed by a bump,
and the narrowest peak with the shortest rise/decay time from 195 to 210 minutes.
The AO system did not work at $t = 131$-$145$ minutes; 
therefore, if there was a flare around 130 minutes, we missed that. 
In the first flare, substructures might exist around 60 minutes.
The rise/decay timescale of $\sim 6.5$ minutes in the third flare
is consistent with the light crossing timescales for 
the inner part of the accretion disk, less than 10 Schwarzschild radii,
around a $4 \times 10^6 M_{\sun}$ black hole.

The middle and bottom panels in Fig. \ref{fig:LCSgrASStars} represent
time evolutions of $Q/I$ and $U/I$, respectively, 
for Sgr A* plus S17 (black line) and the non-variable faint sources (colored lines).
The colored lines show nearly constant values.  
The variation of $Q/I$ for Sgr A* is much larger than those of the other sources.
In the equatorial coordinates, the variation in $U/I$ of SgrA* is 
smaller than $Q/I$.
The error bars of $Q/I$ and $U/I$ for Sgr A* were determined
from the mean RMS of $Q/I$ and $U/I$ for the non-variable faint sources
of similar flux densities.

\begin{figure}[h]
 \begin{center}
  \epsscale{1.0}
   \plotone{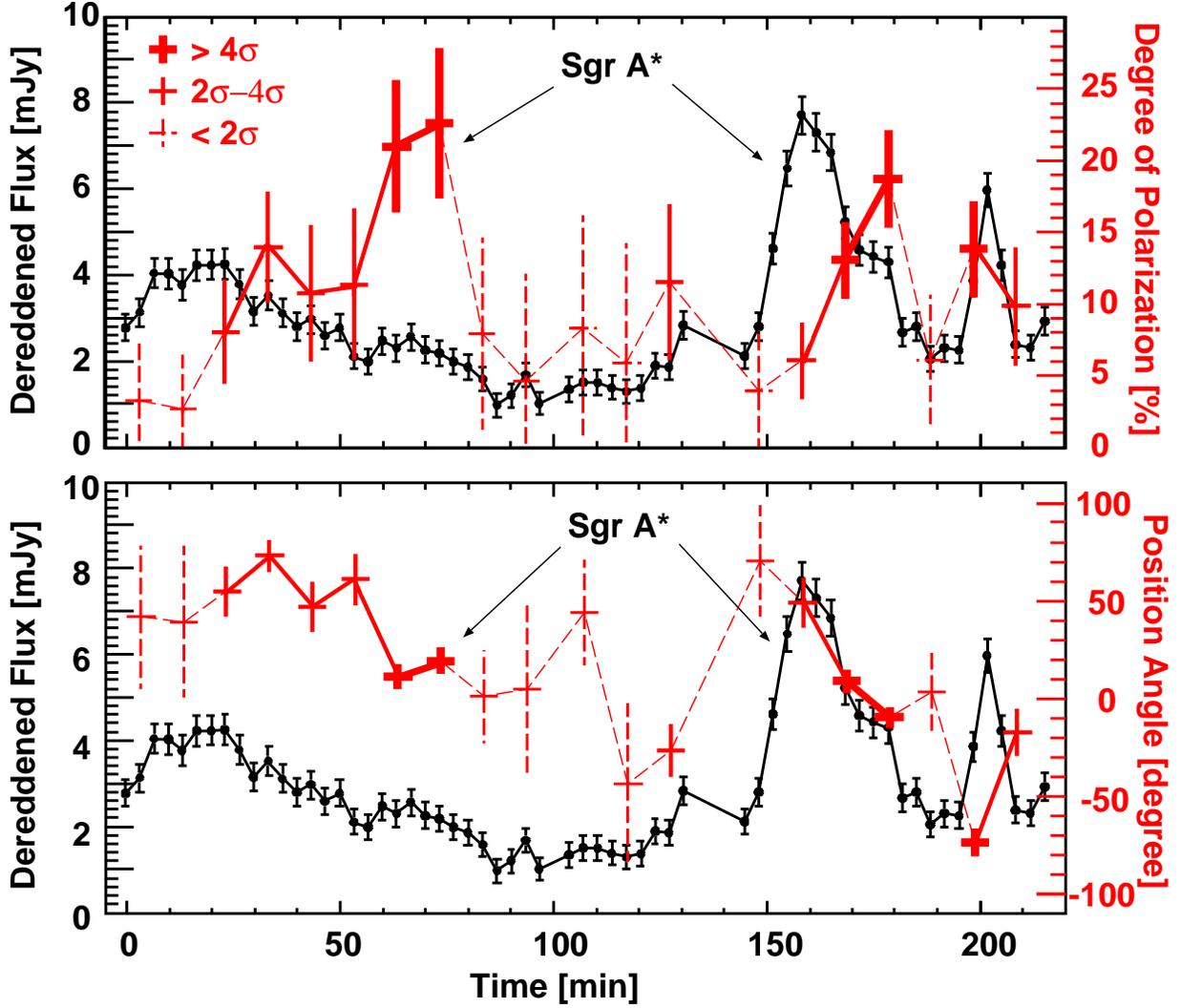}
   \caption{
     Time evolutions of flux and degree of polarization (top),
     and flux and position angle (bottom) for Sgr A*.
     The evolution of flux is shown by black lines, 
     and those of degree of polarization and position angle are shown by red lines.
     Contributions of S17 and background are subtracted in flux.
     The position angle is measured from the north and increasing counterclockwise.
     The data points with $P_{\mathrm{Sgr A^*}}/\sigma_{P_{\mathrm{Sgr A^*}}}<2$
     are shown by crosses with broken red lines,
     and those with $2<P_{\mathrm{Sgr A^*}}/\sigma_{P_{\mathrm{Sgr A^*}}}<4$
     and $P_{\mathrm{Sgr A^*}}/\sigma_{P_{\mathrm{Sgr A^*}}}>4$ are shown by
     crosses with thin and thick red lines, respectively,
     where $P_{\mathrm{Sgr A^*}}$ is the intrinsic degree of polarization of Sgr A*,
     and $\sigma_{P_{\mathrm{Sgr A^*}}}$ is the error of $P_{\mathrm{Sgr A^*}}$.
   }
  \label{fig:PPASgrAS}
 \end{center}
\end{figure}

We find clear variations in the intrinsic degree of polarization $P_{\mathrm{Sgr A^*}}$
and the position angle $\theta_{\mathrm{Sgr A^*}}$ of Sgr A* (Fig. \ref{fig:PPASgrAS}).
The polarization was often detected with not more than $2\sigma$,
and thus we will focus our discussion on the observations
where $P_{\mathrm{Sgr A^*}}$ was detected with more than $2\sigma$ (thin red lines in Fig. \ref{fig:PPASgrAS})
and more than $4\sigma$ (thick red lines) levels.
The first and second flares consist of a weakly or non- polarized main flare and a highly polarized sub-flare.
The $P_{\mathrm{Sgr A^*}}$ rises slowly up to $\sim 22$ \% and probably decays sharply in the first flare.
In the second flare, $P_{\mathrm{Sgr A^*}}$ rises again up to $\sim 19$ \%.
Both rises occur after the bright flare phase, when flux of Sgr A* is decaying.
These profiles are consistent with the previous observations \citep{Meyer06a,Trippe07,Eckart08}.
The variation in $\theta_{\mathrm{Sgr A^*}}$ is complex:
$\theta_{\mathrm{Sgr A^*}}$ swings from $\sim 70\degr$ to $\sim 10\degr$ in the first flare.
The behaviour of $\theta_{\mathrm{Sgr A^*}}$ between the first and second flares is
not clear due to large uncertainties, but
$\theta_{\mathrm{Sgr A^*}}$ swings from $\sim 70\degr$ to $\sim -10\degr$ in the second flare. 
Due to the short duration of only $\sim 13$ minutes,
binning of the data points makes it impossible to trace the variation of
$P_{\mathrm{Sgr A^*}}$ and $\theta_{\mathrm{Sgr A^*}}$ in the third flare.

\section{Discussion}
\label{sec:Discuss}

So far, the time evolutions of polarization of Sgr A* in the NIR bands
have been obtained just for the four flares 
on 
2004 June 12, 
2005 July 29, 
2006 May  31, 
and 2007 May 15 
\citep{Eckart06,Meyer06a,Meyer06b,Trippe07,Eckart08,Zamaninasab08}.
As pointed out by 
\citet{Trippe07}, the observed polarization parameters, 
in particular the position angle, 
in the first three flares show ``remarkable permanence'', 
and the last 2007 flare has also similar parameters in spite of 
some differences \citep{Eckart08}.  
In the typical flare in 2006, 
the degree of polarization rises up to 30\% - 40\%, and simultaneously, 
the position angle swings about $70\degr$
in the decay phase of the broad underlying flare \citep{Meyer06a, Trippe07}.

The first flare we observed shows similar properties to those 
detected in the previous observations.   
It shows a rise in $P_{\mathrm{Sgr A^*}}$ and a swing in $\theta_{\mathrm{Sgr A^*}}$
at $t \ga 50$ minutes, where the long flare (duration of more than 90 minutes) 
is slowly fading away.  
The observed $\theta_{\mathrm{Sgr A^*}}$ during the flare peak is $\sim 60\degr$,
which agrees well with $60\pm20\degr$ of the flare measured by \citet{Eckart06},
and $80\pm10\degr$ by \citet{Meyer06a}.

Such flares can be explained with a hotspot model.  
\citet{Eckart06} made fitting for the flares in 2004 and 2005
with a two-component (hotspot plus disk) model.
For the flares in 2005 and 2006, 
\citet{Meyer06a,Meyer06b} adopted a hotspot plus ring model,
which leads to reasonable fits of polarimetric light curves.
In their models, the broad overall flare is caused by a time varying underlying disk or ring,
and the shorter sub-flares are due to a hotspot on a relativistic orbit around the SMBH.
\citet{Eckart08} proposed a temporary disk with a short jet model,
in which quasi-periodic variation is due to a hotspot on the disk.

The second flare from 145 to 190 minutes seems to be different from other polarized flares.
The flare survives only for $\sim$ 30 minutes,
and is not superimposed on a broader underlying flare.
However, similarly to the previous flares, the second flare shows sub-structures; 
the highest peak is followed by a lower peak or a bump structure.
The two peaks are separate in time by about 15 minutes,
which is very similar to a quasi-periodicity of $15.5\pm2$ minutes obtained by \citet{Meyer06b},
so these sub-structures can be explained by periodic orbital motions of a single hotspot.
If we consider the time evolution of degree of polarization, however, 
we can find similarities between our results and 
the expected light curves for a single orbital motion of a hotspot \citep{Broderick06a}.

Many authors calculated light curves from an emitting bright spot
on the surface of an accretion disk and comoving with the disk
\citep[e.g., ][]{Pineault81,Asaoka89,Bao92,Fukue03}.
Recently, \citet{Broderick06a} calculated light curves in infrared wavelengths, 
including polarization, associated with
a hotspot orbiting the SMBH in the Galactic center.
The primary feature of the light curves 
with a large orbital inclination angle $i$ (close to the edge-on view)
is a narrow and higher peak followed by a broad lower peak/bump.
The first peak is formed by a gravitational lensing effect
which is strongest when the hotspot is right behind the black hole.
Doppler effect and beaming due to the relativistic motion of the hotspot 
in the approaching regime make the second peak/bump.
The time evolutions of {\it polarized} flux show a double-peak profile
with a higher second peak than the first one,
or a slow-rise and sharp-decay profile (with a lower time resolution).
The variation of the position angle is mainly influenced by 
the inclination angle $i$.

The similarities between the second flare and the calculations by \citet{Broderick06a}
suggest that the second flare could be explained with a single orbital motion of a hotspot. 
In the light curve of the second flare from 145 to 190 minutes,
we can see the ``first peak and second peak/bump'' profile.
In the time evolution of degree of polarization,
it is clearly seen that $P_{\mathrm{Sgr A^*}}$ increases in the decay phase of the flare.
The profiles of $P_{\mathrm{Sgr A^*}}$ are asymmetric, showing slow rise and sharp decay.
If the observed time evolution of polarization comes from such a hotspot,
comparison with calculations allows us
to investigate the inclination angle $i$ of the hotspot orbits around Sgr A*,
because the time evolution profile of the flux and the degree of polarization
strongly depend on the inclination angle.
\citet{Broderick06a} showed that when $i \geq 67\fdg5$, 
light curves show a bump after the first peak.
In addition, the time evolution of polarization has a double-peak or 
slow-rise and sharp-decay profile.
These are very similar to the second flare in our observations.
When $i \leq 45\degr$, by contrast, the light curve has a smooth decay phase
and the polarization shows a symmetric or slow-decay profile.
These comparisons could exclude a small inclination angle (near face-on) of 
$i \leq 45\degr$.
This is consistent with recent results,
$i \ga 20\degr$ by \citet{Meyer06a}, 
$i \ga 35\degr$ by \citet{Meyer06b}, 
$i \ga 50\degr$ by \citet{Meyer07},
and $i = 70\degr$ with which \citet{Eckart08} found a minimum reduced-$\chi^2$ value
in the modeling of the observed time evolution of flux and polarization.

Although the hotspot model is a favorite model with our results,
it is premature to use such a simple model to draw strong conclusions.
For example, \citet{Trippe07} proposed that the swing of the position angle in NIR polarization is caused by
either a magnetic field geometry changes due to a vanishing of the accretion disk,
or materials move out of the accretion disk, perhaps into a jet.
A model of the expansion of hot self-absorbed synchrotron plasma blob
was also proposed from multi-wavelengths observations \citep[e.g.,][]{Yusef06,Yusef08,Marrone08}.
This model explains the time delay between different wavelengths in flare emissions,
while the hotspot model does not.
\citet{Meyer08} and \citet{Do09} reported 
non-detection of a statistically significant periodicity in NIR light curves.
Based on the model to a two-temperature magnetorotational instability driven accretion flow by \citet{Liu07},
\citet{Huang08} showed the spectrum and frequency-dependent polarization for Sgr A* with general relativistic effects,
and explain the $90\degr$ flip of the position angle between submillimeter and NIR observations.

Although time-resolved light curves have been presented by many authors 
\citep[e.g.,][]{Goldston05,Falanga07}, 
no ``time-resolved, polarized'' light curves have been available
for the models other than those by \citet{Broderick06a}.  
Polarization provides new information which is extremely useful
to break degeneracy of various model parameters.
Our observations demonstrated that it is now possible to monitor the polarimetric variation of 
Sgr A* continuously up to $\sim 10$ hr by combining contiguous Subaru and VLT observations,
as done for the NIR flux density using VLT and Keck observations \citep{Meyer08}.  
Simulations including polarization evolution to test the various models for Sgr A* 
are now strongly encouraged.

\acknowledgements
This work was supported by Grant-in-Aid for the Global COE Program 
"The Next Generation of Physics, Spun from Universality and Emergence" 
from the Ministry of Education, Culture, Sports, Science and Technology (MEXT) of Japan.


\newpage

\newpage

\end{document}